\documentclass{ifacconf}

\usepackage{graphicx}
\usepackage{natbib}
\usepackage{amsmath}
\usepackage{amssymb}
\usepackage{booktabs}
\usepackage{url}

\newcommand{\xb}{\mathbf{x}}
\newcommand{\xbd}{\dot{\mathbf{x}}}
\newcommand{\xbt}{\mathbf{x}(t)}
\newcommand{\xbh}{\hat{\mathbf{x}}}
\newcommand{\ub}{\mathbf{u}}
\newcommand{\ubt}{\mathbf{u}(t)}
\newcommand{\yb}{\mathbf{y}}
\newcommand{\ybt}{\mathbf{y}(t)}
\newcommand{\ybh}{\hat{\mathbf{y}}}
\newcommand{\jb}{\mathbf{J}}
\newcommand{\jbx}{\mathbf{J}(\xb)}
\newcommand{\jbth}{\mathbf{J}_{\theta}}
\newcommand{\rb}{\mathbf{R}}
\newcommand{\rbx}{\mathbf{R}(\xb)}
\newcommand{\rbth}{\mathbf{R}_{\theta}}
\newcommand{\gb}{\mathbf{G}}
\newcommand{\gbx}{\mathbf{G}(\xb)}
\newcommand{\gbth}{\mathbf{G}_{\theta}}
\newcommand{\hx}{H(\xb)}
\newcommand{\wb}{\mathbf{W}}

\begin{document}
\begin{frontmatter}

\title{Direct Gradient Modelling for Efficient Port-Hamiltonian Neural Networks\thanksref{footnoteinfo1}\thanksref{footnoteinfo2}}

\thanks[footnoteinfo1]{This work is supported in part by the Aalto University House of AI and in part by ABB Oy.}
\thanks[footnoteinfo2]{This work is funded by the European Union (Horizon Europe, ERC, COMPLETE, 101075836). Views and opinions expressed are however those of the author(s) only and do not necessarily reflect those of the European Union or the European Research Council Executive Agency. Neither the European Union nor the granting authority can be held responsible for them.}

\author[First]{Jussi Särkilahti}
\author[Second]{Gé Jan Ember van Otterdijk}
\author[First]{Sadegh Ebrahimkhani}
\author[First]{Junyi Li}
\author[First]{Marko Hinkkanen}
\author[Second]{Maarten Schoukens}

\address[First]{Aalto University, Espoo, Finland (e-mail: j.p.sarkilahti@aalto.fi)}
\address[Second]{Eindhoven University of Technology, Eindhoven, The Netherlands (e-mail: g.j.e.v.otterdijk@tue.nl)}

\begin{abstract}
Port-Hamiltonian neural networks learn passive models of physical systems directly from input-output data. Their dynamics depend on the Hamiltonian, which describes the system's stored energy, only through its gradient. Existing methods obtain the gradient via automatic differentiation of a neural Hamiltonian function. This paper utilizes a gradient network to directly model the Hamiltonian gradient. The resulting model remains port-Hamiltonian while omitting the differentiation step. The effectiveness of direct gradient modelling is tested against the automatic differentiation method on a simulated mass-spring-damper system and the real-world cascaded tanks benchmark. The proposed method attains better accuracy while using fewer parameters and achieving a more reliable learning procedure.
\end{abstract}


\end{frontmatter}

{\renewcommand{\thefootnote}{}\footnotetext{This work has been submitted to IFAC for possible publication.}}

\section{Introduction}
\label{sec:introduction}
Accurate models of dynamical systems are needed for simulation, analysis, and control. Deriving such models from physical laws can be difficult and time-consuming, particularly for nonlinear systems. Additionally, in fields where the dynamics are not fully understood, the first-principles approach may lead to structural modelling errors \citep{ma2019first, nieminen2009issues}.

Nonlinear system identification methods estimate the models directly from measured data \citep{schoukens2019nonlinear}. Methods like state-space neural networks were first introduced in the 1990s \citep{suykens1996artificial}, and have seen continued interest until recently \citep{masti2021learning, beintema2023deep}. These black-box models are system-agnostic and therefore applicable to a wide range of identification tasks. They are, however, hard to interpret, and they may extrapolate poorly outside the measured operating range \citep{wang2021bridging}.

Physics-informed methods combine physical knowledge with data-driven learning \citep{wang2025physics}. For example, a \emph{Hamiltonian neural network} (HNN) learns the Hamiltonian, e.g., the energy function of a conservative system \citep{greydanus2019hamiltonian}. The HNN trains faster and generalizes better than a black-box neural network model, but it cannot model energy dissipation or external inputs. \emph{Port-Hamiltonian} (PH) systems include both \citep{vanderschaft2014porthamiltonian}. Furthermore, PH models are passive, can be interconnected, and are applicable across multiple physical domains. Several works combine neural networks with the PH structure. \citet{desai2021port} add a state-independent dissipation term and a time-dependent external force to the HNN, while \citet{cherifi2026nonlinear} identify general nonlinear PH systems from input, state, and output data. The \emph{output-error PH neural network} (OE-PHNN) of \citep{moradi2026port} needs only input and output data. It handles output measurement noise and its estimates are proven to be consistent.

As training an OE-PHNN is a nonconvex optimization problem, the learning procedure can get stuck in poor local minima. This issue is addressed in part in \citep{vanotterdijk2026improved}, where a linear estimate of the port-Hamiltonian system is used to give the OE-PHNN a better parameter initialization. Nevertheless, further opportunities remain in the parametrization of the model. This paper considers the direct modelling of the Hamiltonian's gradient to further improve the learning procedure.
In conventional PH neural networks, the model’s system matrices are constrained to the PH structure while the Hamiltonian is represented by an unconstrained, scalar-output neural network. The dynamics need only the gradient of the Hamiltonian, which is computed by automatic differentiation. Backpropagation through these dynamics requires mixed second-order derivatives of the Hamiltonian network with respect to its inputs and parameters, which introduces additional computational overhead. \cite{roth2025stable} constrain the Hamiltonian to be convex, by using an \emph{input convex neural network} \citep{amos2017input}, but still compute its gradient by automatic differentiation.

This paper models the Hamiltonian gradient directly with a \emph{gradient network} \citep{chaudhari2024gradient}. The output of a gradient network is, by construction, the gradient of a scalar function, making it suitable for Hamiltonian gradient modelling. Additionally, in a monotone gradient network, this scalar function is convex. Previously, gradient networks have been used to learn Hamiltonian dynamics \citep{chaudhari2024gradient} and magnetic models of electric machines \citep{li2026gradient}.

The contributions of this paper are:
\begin{itemize}
    \item The Hamiltonian gradient of the OE-PHNN is modelled directly using a monotone gradient network. The resulting model preserves the PH structure, has a convex Hamiltonian bounded from below, and eliminates the automatic differentiation during prediction.
    \item The proposed method is compared with the auto\-matic-differentiation-based OE-PHNN on a simulated mass-spring-damper system and the real-world cascaded tanks benchmark.
\end{itemize}

With regard to the rest of the paper, Section~\ref{sec:preliminaries} reviews PH systems, the OE-PHNN, and gradient networks. Section~\ref{sec:methods} presents the proposed model. Section~\ref{sec:results} reports the numerical experiments, while Section~\ref{sec:conclusion} provides a conclusion.

\section{Preliminaries}
\label{sec:preliminaries}

\subsection{Port-Hamiltonian Systems}
\label{sec:PHS}
A PH system without feedthrough can be described by
\begin{subequations}
    \label{eq:PHS}
    \begin{alignat}{2}
        \xbd(t) &= \bigl[\jb(\xbt)\!-\!\rb(\xbt)\bigr]\nabla H(\xbt)
        \!+\!\gb(\xbt)\ubt\mkern-2mu, \\
        \ybt &= \gb^{\top}(\xbt)\nabla H(\xbt)
    \end{alignat}
\end{subequations}
where $\xbt \in \mathbb{R}^{n_{x}}$ is the state, $\ubt \in \mathbb{R}^{n_{p}}$ is the input, and $\ybt \in \mathbb{R}^{n_{p}}$ is the output. The time argument is hereafter omitted for brevity. The scalar function $\hx$ is the Hamiltonian, i.e., the energy stored in the system. The skew-symmetric matrix $\jbx \in \mathbb{R}^{n_{x} \times n_{x}}$ describes the internal energy exchange. The symmetric positive semidefinite matrix $\rbx \in \mathbb{R}^{n_{x} \times n_{x}}$ describes the energy dissipation. The port matrix $\gbx \in \mathbb{R}^{n_{x} \times n_{p}}$ describes how the inputs act on the system and how the Hamiltonian gradient is mapped to the outputs.

Since $\jb$ is skew-symmetric and $\rb$ is positive semidefinite, \eqref{eq:PHS} gives
\begin{equation}
    \label{eq:passivity}
    \dot{H} = -\nabla H^{\top} \rb \nabla H + \yb^{\top}\ub \leq \yb^{\top}\ub,
\end{equation}
where $\dot{H}$ is the time derivative of the Hamiltonian. Thus, the stored energy can increase only through power $\yb^{\top}\ub$ supplied via the input ports. If $H$ is bounded from below, the system is also passive.

The input and the output are sampled with the sampling interval $T_\mathrm{s}$. The samples are $\ub_{k} = \ub(kT_\mathrm{s})$ and $\yb_{k} = \yb(kT_\mathrm{s}) + \varepsilon_{k}$, where $\varepsilon_{k}$ is zero-mean measurement noise with finite variance. The resulting dataset is
\begin{equation}
    \mathcal{D}_N=\{(\ub_{k}, \yb_{k})\}_{k=0}^{N-1}.
    \label{eq:dataset}
\end{equation}
The task is to identify a model of \eqref{eq:PHS} using this dataset. The state is not measured.

\subsection{Output-Error Port-Hamiltonian Neural Networks}
\label{sec:PHNN}
The OE-PHNN \citep{moradi2026port} uses the SUBNET method \citep{beintema2023deep}. The data are divided into overlapping sections of $T$ samples. Each section starting at time index $t$ is simulated in three steps. First, an encoder $\psi_{\theta}$ estimates the initial state from the $n_{a}$ previous outputs and the $n_{b}$ previous inputs. Second, an ordinary differential equation (ODE) solver simulates the state forward in time. Third, the outputs are computed from the simulated states. The parameters $\theta$ of all parts are estimated together by minimizing the simulation error \eqref{subeq:loss_theta} subject to \eqref{subeq:encoder}--\eqref{subeq:decoder}:
\begin{subequations}
\label{eq:SUBNET}
\begin{align}
    V(\theta) &= \frac{1}{C} \sum_{t=n}^{N-T}\sum_{k=0}^{T-1} \left\|\yb_{t+k}- \ybh_{t+k|t}\right\|_2^2 \label{subeq:loss_theta}\\
    \xbh_{t|t} &= \psi_\theta (\yb_{t-n_{a}}^{t-1}, \ub_{t-n_{b}}^{t-1}) \label{subeq:encoder}\\
    \xbh_{t+k+1|t} &= \text{ODE-solver} [f_{\theta}(\xbh_{t+k|t}, \ub_{t+k})] \label{subeq:state_propagation}\\
    \ybh_{t+k|t} &= h_{\theta}(\xbh_{t+k|t})    \label{subeq:decoder}
\end{align}
\end{subequations}
where $\ybh_{t+k|t}$ is the output at time $t+k$ simulated from the start time $t$, $n = \max(n_a, n_b)$, and $C = (N-T-n+1)T$ is the number of terms in the sum. The encoder input $\ub_{t-n_{b}}^{t-1}$ contains the inputs $\ub_{t-n_{b}}, \ldots, \ub_{t-1}$, and $\yb_{t-n_{a}}^{t-1}$ contains the outputs $\yb_{t-n_{a}}, \ldots, \yb_{t-1}$. The ODE solver integrates the state equation over one sampling interval, with the input held constant. A longer horizon $T$ increases the computational cost of training.

In the OE-PHNN, the state equation and the output equation have the PH form:
\begin{subequations}
    \label{eq:OE-PHNN}
    \begin{align}
        f_{\theta}(\xbh, \ub) &= \bigl[ \jbth(\xbh)-\rbth(\xbh)\bigr]\nabla H_{\theta}(\xbh) + \gbth(\xbh)\ub \label{eq:state_evol_param} \\
        h_{\theta}(\xbh) &= \gbth^{\top}(\xbh)\nabla H_{\theta}(\xbh). \label{eq:output_eq_param}
    \end{align}
\end{subequations}
The system matrices are constructed from the outputs of multilayer perceptrons (MLPs), i.e., standard feedforward neural networks. The skew-symmetric matrix is $\jbth = \mathbf{A}_{\theta} - \mathbf{A}^{\top}_{\theta}$, and the positive semidefinite matrix is $\rbth = \mathbf{B}_{\theta}\mathbf{B}^{\top}_{\theta}$, where $\mathbf{A}_{\theta}(\xbh)$ and $\mathbf{B}_{\theta}(\xbh)$ are MLP outputs. The matrix $\gbth(\xbh)$ is the output of a third MLP. In the OE-PHNN of \citep{moradi2026port} as well as in the automatic-differentiation-based model used as comparison in this work, the Hamiltonian $H_{\theta}$ is a scalar MLP whose output passes through an exponential linear unit (ELU). The ELU keeps $H_{\theta}$ bounded from below. The gradient $\nabla H_{\theta}$ is then computed via automatic differentiation.

\begin{figure*}[t]
    \centering
    \includegraphics[width=.94\textwidth]{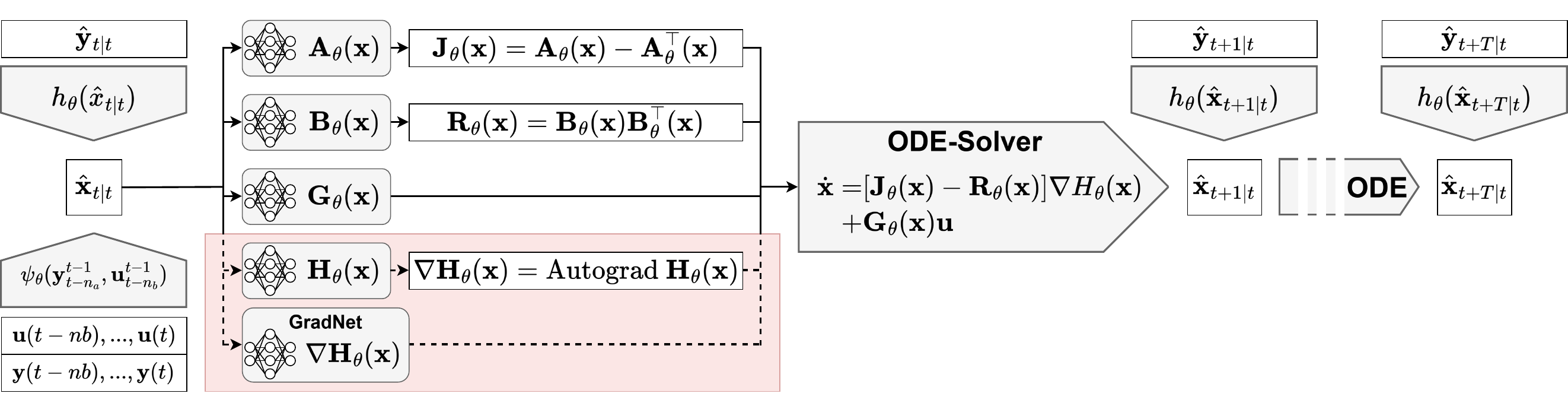}
    \caption{Model structure. The red area shows the two compared ways to obtain the Hamiltonian gradient: an MLP for the Hamiltonian followed by automatic differentiation, or a gradient network which directly gives the gradient.}
    \label{fig:structure-change}
\end{figure*}

\subsection{Gradient Networks}
\label{subsec:gradient_networks}
A \emph{gradient network} (\texttt{GradNet}) is a neural network whose output is, by construction, the gradient of a scalar function \citep{chaudhari2024gradient}. A simple example is
\begin{equation}
    \label{eq:GradNet}
    \mathbf{g}(\xb) = \wb^{\top}\sigma(\wb \xb + \mathbf{b}) + \mathbf{a}
\end{equation}
where $\wb$ is a weight matrix, $\mathbf{a}$ and $\mathbf{b}$ are bias vectors, and the activation function $\sigma$ is applied elementwise. If $\sigma = \phi'$ for a scalar function $\phi$, then \eqref{eq:GradNet} is the gradient of
\begin{equation}
    \label{eq:GradNet_potential}
    H(\xb) = \sum_{i} \phi(\mathbf{w}_{i}^{\top}\xb + b_{i}) + \mathbf{a}^{\top}\xb
\end{equation}
where $\mathbf{w}_{i}^{\top}$ is the $i$th row of $\wb$ and $b_{i}$ is the $i$th element of $\mathbf{b}$. Hence, the network gives the gradient $\nabla H$ without computing $H$ itself. If $\phi$ is also convex, then $\sigma$ is nondecreasing ($\sigma' \geq 0$), the Hessian $\nabla^{2} H = \wb^{\top}\mathrm{diag}(\sigma')\wb$ is positive semidefinite, and $H$ is convex. Such networks are called \emph{monotone gradient networks} (\texttt{mGradNets}). Note that a convex function is not necessarily bounded from below. For example, the linear term $\mathbf{a}^{\top}\xb$ in \eqref{eq:GradNet_potential} is convex but unbounded.

More expressive networks are obtained by summing several modules. The \emph{modular monotone gradient network} (\texttt{mGradNet-M}) \citep{chaudhari2024gradient} is
\begin{equation}
\label{eq:GradNet-M}
    \mathbf{g}(\xb) = \mathbf{a} + \sum^{M}_{m=1} \rho_{m}(\phi_{m}(\mathbf{z}_{m}))\wb^{\top}_{m}\sigma_{m}(\mathbf{z}_{m})
\end{equation}
where $\mathbf{z}_{m} = \wb_{m}\xb + \mathbf{b}_{m}$ and $M$ is the number of modules. Each module has a weight matrix $\wb_{m}$, a bias vector $\mathbf{b}_{m}$, a convex scalar function $\phi_{m}$ of the vector $\mathbf{z}_{m}$, its gradient $\sigma_{m} = \nabla \phi_{m}$, and a nonnegative nondecreasing scalar function $\rho_{m}$. By the chain rule, the $m$th term in \eqref{eq:GradNet-M} is the gradient of $S_{m}(\phi_{m}(\mathbf{z}_{m}))$, where $S_{m}' = \rho_{m}$. This function is convex, because $S_{m}$ is convex and nondecreasing and $\phi_{m}(\mathbf{z}_{m})$ is convex in $\xb$. Hence, \eqref{eq:GradNet-M} is the gradient of a convex function. \citet{chaudhari2024gradient} show that monotone gradient networks can approximate gradients of convex functions arbitrarily well.

\section{Proposed Model}
\label{sec:methods}
The OE-PHNN \eqref{eq:OE-PHNN} depends on the Hamiltonian only through its gradient. This paper therefore models the gradient directly:
\begin{equation}
    \label{eq:proposed}
    \nabla H_{\theta}(\xbh) = \mathbf{g}_{\theta}(\xbh) + \mu^{+}\xbh,
\end{equation}
where $\mathbf{g}_{\theta}$ is the \texttt{mGradNet-M} \eqref{eq:GradNet-M}, with modules defined by
\begin{align}
    \phi_m(\mathbf{z}_m)
    &= \sum_i \frac{\lambda_{m,i}^{+}}{B_{m,i}^{+}}
    \operatorname{softplus}(B_{m,i}^{+}z_{m,i}) + c_m, \\
    \sigma_m(\mathbf{z}_m) &= \nabla\phi_m(\mathbf{z}_m), \\
    \rho_m(s) &= a_m^{+}\operatorname{sigmoid}(s).
\end{align}
Here, $c_m$ is a trainable offset, $\lambda_{m,i}^{+}=\operatorname{softplus}(\lambda_{m,i})$,
$B_{m,i}^{+}=\operatorname{softplus}(B_{m,i})+\varepsilon$,
$a_m^{+}=\operatorname{softplus}(a_m)$, and
$\mu^{+}=\operatorname{softplus}(\mu)+\varepsilon$, with
$\varepsilon=10^{-6}$.
All other parts of the OE-PHNN are unchanged: the encoder, the MLPs for $\jbth$, $\rbth$, and $\gbth$, and the ODE solver. Fig.~\ref{fig:structure-change} illustrates the change in the model structure.

The Hamiltonian $H_{\theta}$ itself is never computed. However, \eqref{eq:proposed} is, by construction, the gradient of a convex function plus the quadratic function $\mu^{+}\|\xbh\|^{2}/2$. As such, $H_{\theta}$ must be a strongly convex function with a unique minimum and is therefore bounded from below. By \eqref{eq:passivity}, the overall model remains passive as it retains the PH structure. The proposed model has the following properties:
\begin{itemize}
    \item The gradient of the Hamiltonian is computed directly in the forward pass, avoiding the automatic differentiation necessary in the original model structure.
    \item The Hamiltonian is convex by construction. This holds for many physical systems, such as the mass-spring-damper system in Section~\ref{sec:results}. Systems with nonconvex Hamiltonians can be represented by the more general \emph{modular gradient network} (\texttt{GradNet-M}) \citep{chaudhari2024gradient}.
    \item The Hamiltonian representation needs fewer parameters (see Table~\ref{tab:settings}).
    \item The proposed \texttt{mGradNet-M} parametrization maintains the port-Hamiltonian structure and thus retains the consistency analysis of \citep{moradi2026port}.
\end{itemize}

\section{Numerical Experiments}
\label{sec:results}
To examine the difference between the discussed model structures, a simulation study is performed. To ensure that the implementations are identical in everything besides the model structure, two models are trained, denoted \texttt{AG-ours} and \texttt{MGM}, which differ only in their parametrization of the Hamiltonian, as \texttt{AG-ours} follows the default parametrization \eqref{eq:OE-PHNN} while \texttt{MGM} follows the model proposed in Section~\ref{sec:methods}. These models are compared to the available reference results \texttt{AG-ref} from \citep{moradi2026port}. Experiments on a simulated mass-spring-damper (MSD) system and the real-world cascaded tanks (CT) benchmark \citep{schoukens2016cascaded} examine the effects of initialization, noise, training data size, and simulation horizon.

\subsection{Test Systems}
\subsubsection{Mass-spring-damper system}
\begin{figure}[t]
    \centering
    \includegraphics[width=.85\linewidth]{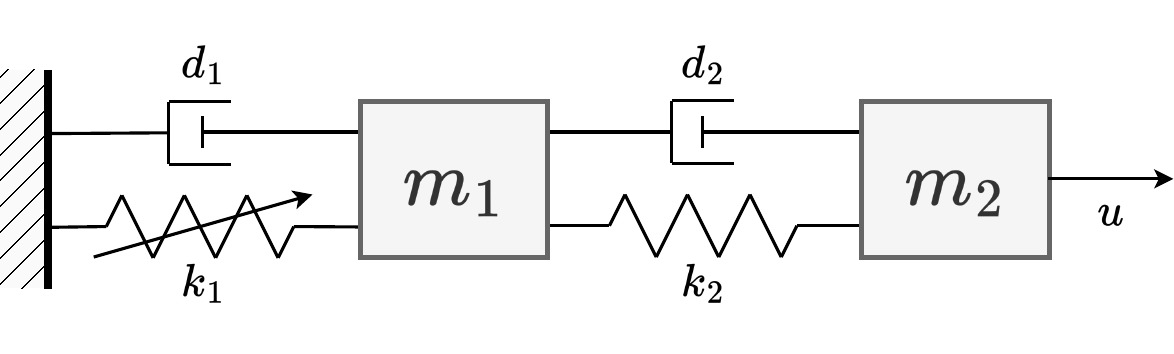}
    \caption{Coupled mass-spring-damper system. The first spring is nonlinear.}
    \label{fig:MSD-setup}
\end{figure}

The MSD system in Fig.~\ref{fig:MSD-setup} is the same as in \citep{moradi2026port}. Its equations of motion are
\begin{equation} \label{eq:3MSD}
    \mathbf{M}\ddot{\mathbf{q}} + \mathbf{D}\dot{\mathbf{q}} + \mathbf{k}(\mathbf{q}) = \begin{bmatrix} 0 \\ 1 \end{bmatrix} u
\end{equation}
where $\mathbf{q} = [q_{1}, q_{2}]^{\top}$ contains the positions of the masses $m_{1}$ and $m_{2}$, $\mathbf{M}$ is the mass matrix, $\mathbf{D}$ is the damping matrix, $\mathbf{k}(\mathbf{q})$ contains the spring forces, and $u$ is the external force acting on the outer mass. The force of the first spring is $k_{1,\mathrm{l}}q_{1} + k_{1,\mathrm{c}}q_{1}^{3}$. The parameters are $m_{1} = m_{2} = 1$~kg, $k_{1,\mathrm{l}} = k_{2} = 1$~N/m, $k_{1,\mathrm{c}} = 0.1$~N/m$^{3}$, and $d_{1} = d_{2} = 0.5$~N\,s/m.

The input is a random-phase multisine with 100 harmonics of the fundamental frequency $f_0=0.01$~Hz, with new phases drawn uniformly from $[0,2\pi)$ and the RMS value scaled to 0.5~N for each 100-s period. The input is held constant between samples. The data are generated with the fourth-order Runge--Kutta (RK4) method using the step $T_\mathrm{s} = 0.1$~s. The output is the velocity $y = \dot{q}_{2}$ of the outer mass. White Gaussian noise is added to the training and validation outputs. The noise level is given by the signal-to-noise ratio $\mathrm{SNR} = 10\log_{10}(P_{y}/P_{\varepsilon})$, where $P_{y}$ and $P_{\varepsilon}$ are the mean powers of the noise-free output and the noise. The default SNR is 50~dB. The test data are noise-free. A 28\,000-sample record is split into 20\,000 training and 8000 validation samples, while an independent 20\,000-sample record is used for testing.

\subsubsection{Cascaded tanks}
The CT system \citep{schoukens2016cascaded} consists of two cascaded tanks with free outlets, with a pump controlling the inflow to the upper tank. The normally weakly nonlinear flow can exhibit hard saturation upon overflow, which occurs in the training, validation, and testing datasets and makes the system challenging to model. The input is pump voltage and the output is the lower-tank water level. Since their product is not physical power, the PH formulation is used only as a model structure. The benchmark contains separate 1024-sample training and test records with $T_\mathrm{s}=4$~s. Following \citep{moradi2026port}, the first 512 test samples are also used for validation, so the test results are not fully independent of model selection. Following \citep{moradi2026port}, the ODE solver uses a step of 0.04 instead of 4~s to prevent numerical overflow during training. This rescales model time by a factor of 100 without changing sampled predictions. For \texttt{AG-ours}, training-set mean and standard-deviation normalization of inputs and outputs gave the best performance, whereas normalization did not improve \texttt{MGM} performance and was omitted.

\begin{table}[t]
    \centering
    \caption{Model and training settings. MLP size $a\times b$: $a$ hidden layers of $b$ neurons, \texttt{mGradNet-M} size $M\times d$: $M$ modules with $\mathbf{z}_m\in\mathbb{R}^d$. Hamiltonian parameter counts are in parentheses, and learning rates are given as epoch:value.}
    \label{tab:settings}
    \setlength{\tabcolsep}{3pt}
    \begin{tabular}{@{}p{0.44\columnwidth}cc@{}}
    \toprule
     & MSD & CT \\
    \midrule
    State dimension $n_{x}$ & 4 & 2 \\
    Encoder MLP & $2 \times 64$ & $2 \times 64$ \\
    Encoder inputs $n_{a} = n_{b}$ & 20 & 20 \\
    MLPs of $\mathbf{A}_{\theta}$, $\mathbf{B}_{\theta}$, and $\gbth$ & $1 \times 8$ & $1 \times 8$ \\
    \texttt{AG-ours}: MLP of $H_{\theta}$ & $2 \times 16$ (369) & $1 \times 8$ (69) \\
    \texttt{MGM}: \texttt{mGradNet-M} & $4 \times 4$ (126) & $2 \times 2$ (28) \\
    Horizon $T$ & 200 & 120 \\
    Batch size & 256 & 32 \\
    Epochs & 70 & 700 \\
    Epoch block boundaries &
    \begin{tabular}[c]{@{}c@{}}
    $10,20,30,$ \\
    $40,50,70$
    \end{tabular} &
    \begin{tabular}[c]{@{}c@{}}
    $10,20,30,50,70,$ \\
    $100,150,200,250,$ \\
    $300,400,500,700$
    \end{tabular} \\
    \texttt{AG-ours} learning rate &
    $0:10^{-3}$ &
    $0:10^{-3}$ \\
    \texttt{MGM} learning rate &
    \begin{tabular}[c]{@{}c@{}}
    $0:10^{-2}$ \\
    $20:5\cdot10^{-3}$ \\
    $30:2\cdot10^{-3}$ \\
    $40:10^{-3}$
    \end{tabular} &
    \begin{tabular}[c]{@{}c@{}}
    $0:10^{-2}$ \\
    $10:5\cdot10^{-3}$ \\
    $30:10^{-3}$ \\
    $600:2\cdot10^{-4}$
    \end{tabular} \\
    RK4 step & 0.1~s & 0.04 \\
    \bottomrule
    \end{tabular}
\end{table}

\subsection{Models and Training}
Table~\ref{tab:settings} lists the model and training settings. The \texttt{AG-ours} settings follow \citep{moradi2026port}. The ODE solver uses one RK4 integration step, and all MLPs use tanh activations.
All models are trained with Adam in epoch blocks specified in Table~\ref{tab:settings}. After each block, the parameters with the lowest validation loss so far are restored and the batches reshuffled. For MSD, the first five epochs use a shorter horizon $T=5$.
The learning-rate schedules were selected based on validation performance. For \texttt{AG-ours}, a fixed learning rate of $10^{-3}$ gave the best results, while \texttt{MGM} benefitted from a higher initial rate followed by stepwise reductions. The full code including the optimization strategy is available at
GitHub: \url{https://github.com/jsarkilahti/direct_gradient_modelling}.

For MSD, validation uses the full-record simulation error. For CT, the validation loss is $0.9L_\mathrm{trunc}+0.1L_\mathrm{full}$, where $L_\mathrm{trunc}$ is \eqref{subeq:loss_theta} and $L_\mathrm{full}$ is the full-record simulation error.
This choice leverages the strong convexity and monotonicity of \texttt{MGM}, which constrain the learned Hamiltonian gradient. The truncated prediction loss guides early training, while the full-record simulation loss promotes long-term predictive accuracy. The parameters with the lowest validation loss are selected for testing. Ten random initializations are used for each system, with both models sharing the initial encoder, $\jbth$, $\rbth$, and $\gbth$ parameters for each initialization.

Testing uses a full-record simulation initialized by the encoder. Accuracy is measured by the \emph{normalized root mean square error} (NRMSE)
\begin{align*}
    \mathrm{NRMSE}
    =
    \frac{
        \sqrt{\frac{1}{N_\mathrm{e}}\sum_{k=1}^{N_\mathrm{e}}
        \left\|y_k-\hat{y}_k\right\|_2^2}
    }{\sigma_y},
\end{align*}
where $y_k$ and $\hat{y}_k$ are the measured and simulated outputs, $\sigma_{y}$ is the standard deviation of the measured output, and $N_\mathrm{e}$ is the number of simulated test samples.

\subsection{Results}
\subsubsection{Random initialization}
Table~\ref{tab:init} shows the test errors over the ten initializations. For MSD, \texttt{MGM} has a lower mean and minimum NRMSE than both \texttt{AG-ours} and \texttt{AG-ref}. However, the minimum for \texttt{MGM} appears to be an outlier as the best validation loss is 29 times lower than the second best, while this ratio is 2.8 for \texttt{AG-ours}. At the same time, the NRMSE of the median models is close to their mean at 0.069 for \texttt{MGM} and 0.102 for \texttt{AG-ours}, showing that the accuracy of the obtained models is relatively consistent. For CT, the \texttt{AG-ours} models are substantially less accurate than \texttt{AG-ref}, obtained with $T=60$ and four RK4 steps per sampling interval. The discrepancy likely reflects differences in training settings and the number of random initializations. The best \texttt{MGM} model still performs slightly better. Again the median NRMSEs are close to the mean, at 0.170 for \texttt{MGM} and 0.725 for \texttt{AG-ours}.

\begin{table}[t]
    \centering
    \caption{Model NRMSE on the MSD and CT test sets over ten random initializations.}
    \label{tab:init}
    \begin{tabular}{lccccc}
    \toprule
    & \texttt{AG-ref} & \multicolumn{2}{c}{\texttt{AG-ours}} & \multicolumn{2}{c}{\texttt{MGM}} \\
    \cmidrule(lr){3-4} \cmidrule(lr){5-6}
     & & Mean & Min & Mean & Min \\
    \midrule
    MSD & 0.019 & 0.097 & 0.063 & 0.063 & \textbf{0.015} \\
    CT & 0.133  & 0.689 & 0.463 & 0.212 & \textbf{0.125} \\
    \bottomrule
    \end{tabular}
\end{table}

\begin{figure}
    \centering
    \includegraphics[width=\linewidth]{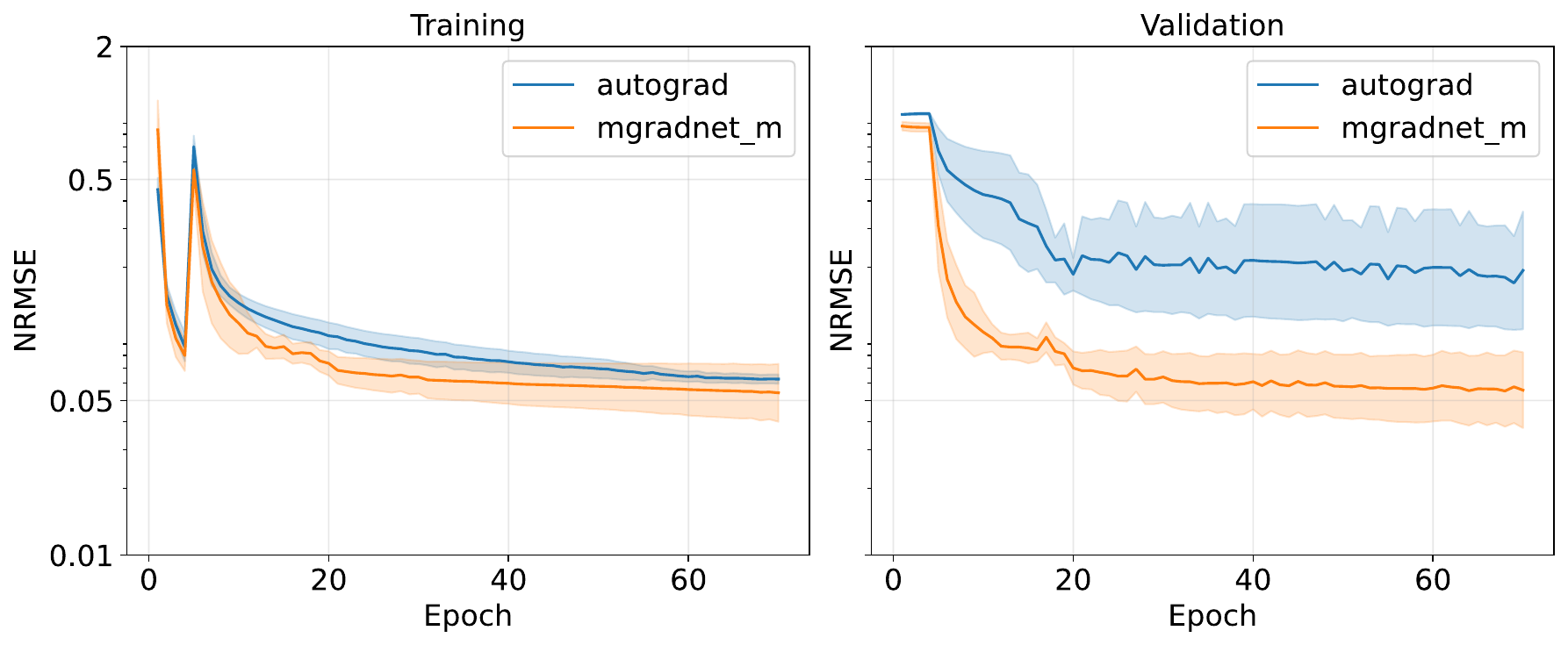}\vspace{-3mm}
    \caption{MSD: geometric mean of training and validation losses (solid line), 95\% confidence interval (shaded).}
    \label{fig:MSD-losses}
\end{figure}

\begin{figure}
    \centering
    \includegraphics[width=\linewidth]{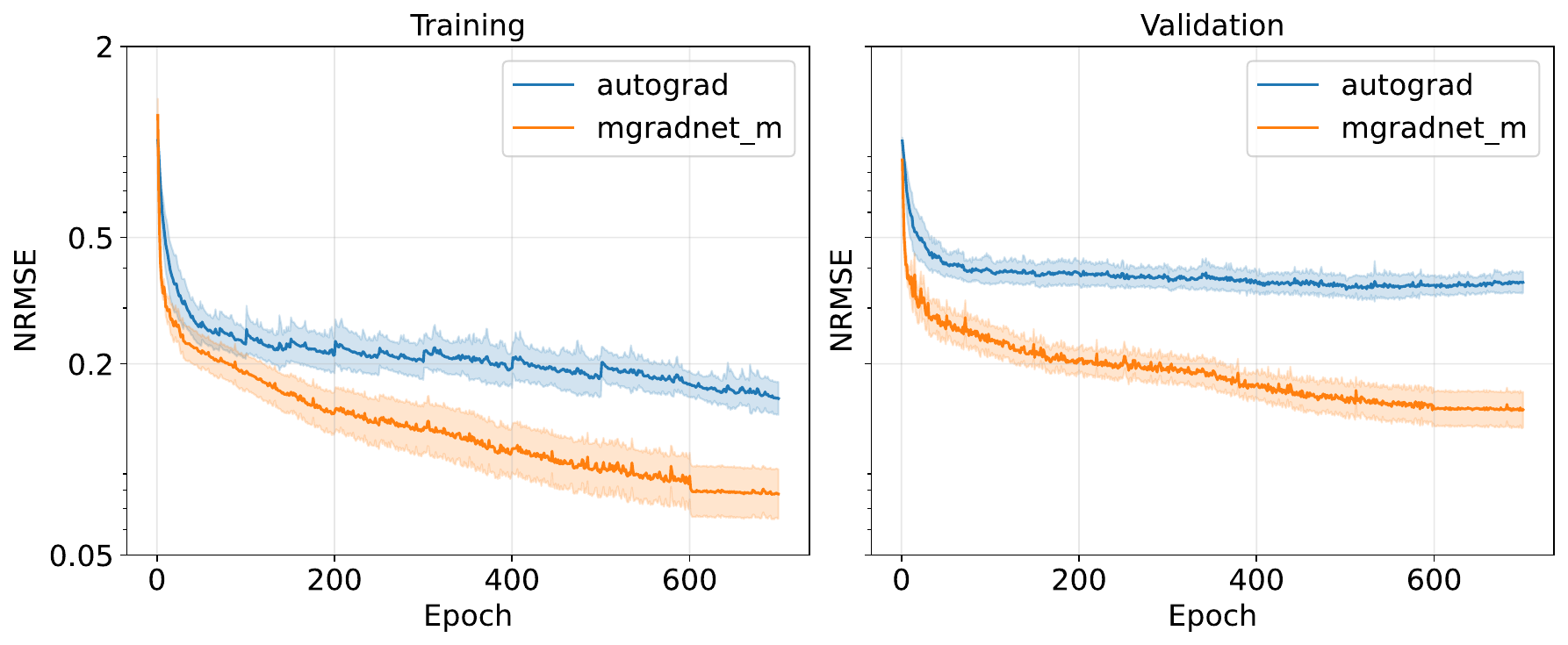}\vspace{-3mm}
    \caption{CT: mean training and validation losses (solid line), 95\% confidence interval (shaded).}
    \label{fig:CT-losses}
\end{figure}

Figs.~\ref{fig:MSD-losses} and \ref{fig:CT-losses} show the geometric mean of training and validation NRMSE on logarithmic axes with 95\% confidence intervals.
For MSD, the validation losses of \texttt{AG-ours} vary widely between the initializations, and two runs deteriorate during training. Nine of the ten \texttt{MGM} runs converge to similar losses, indicating better robustness to initialization in this case. The spike at epoch 5 is caused by the change from $T = 5$ to $T = 200$. For CT, both models train smoothly. The validation loss of \texttt{MGM} remains consistently below that of \texttt{AG-ours} and continues to decrease until the end of training, while the validation loss of \texttt{AG-ours} levels off early, illustrating the improvement in model quality obtained by the direct gradient parametrization.

\begin{figure}
    \centering
    \includegraphics[width=.9\linewidth]{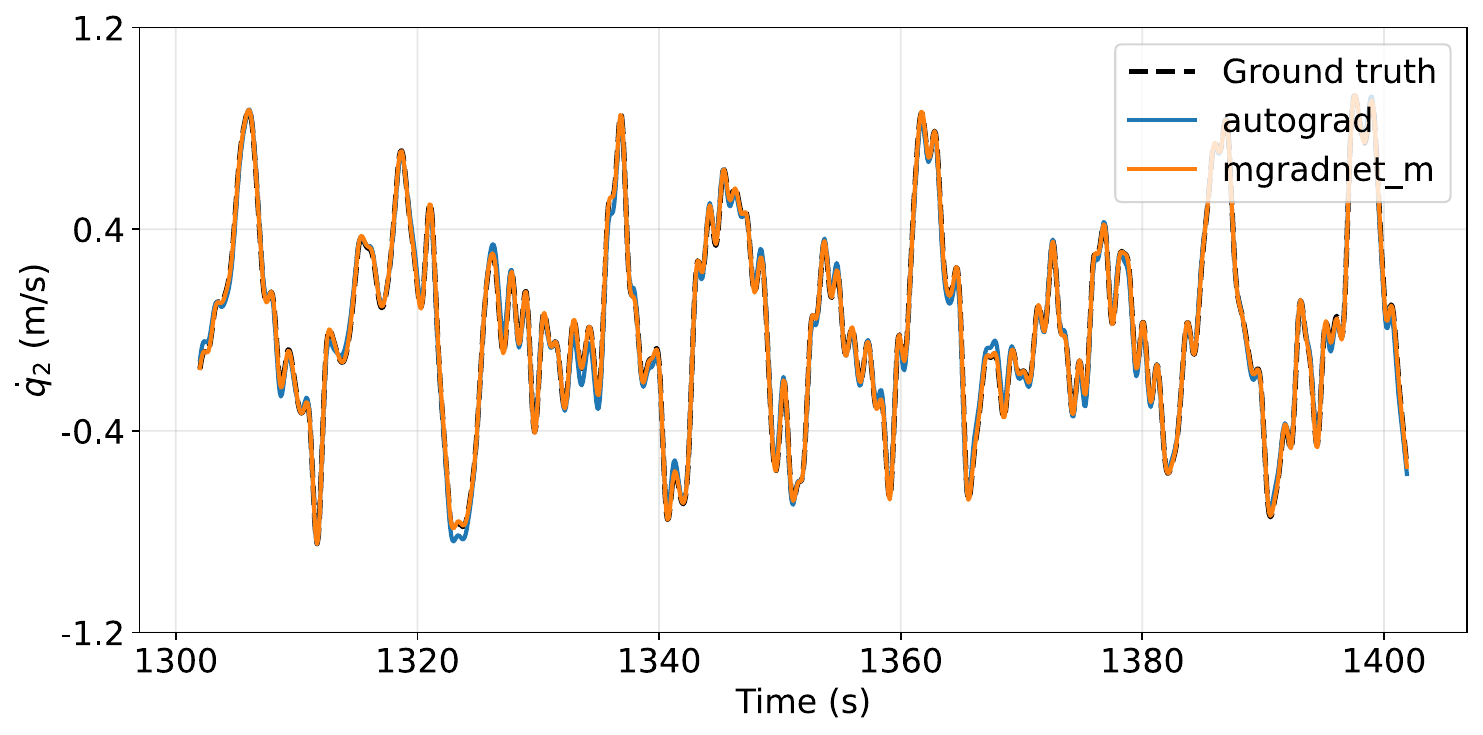}\vspace{-3mm}
    \caption{MSD: test simulation, 100-s section.}
    \label{fig:MSD-simulation}
\end{figure}

\begin{figure}
    \centering
    \includegraphics[width=.9\linewidth]{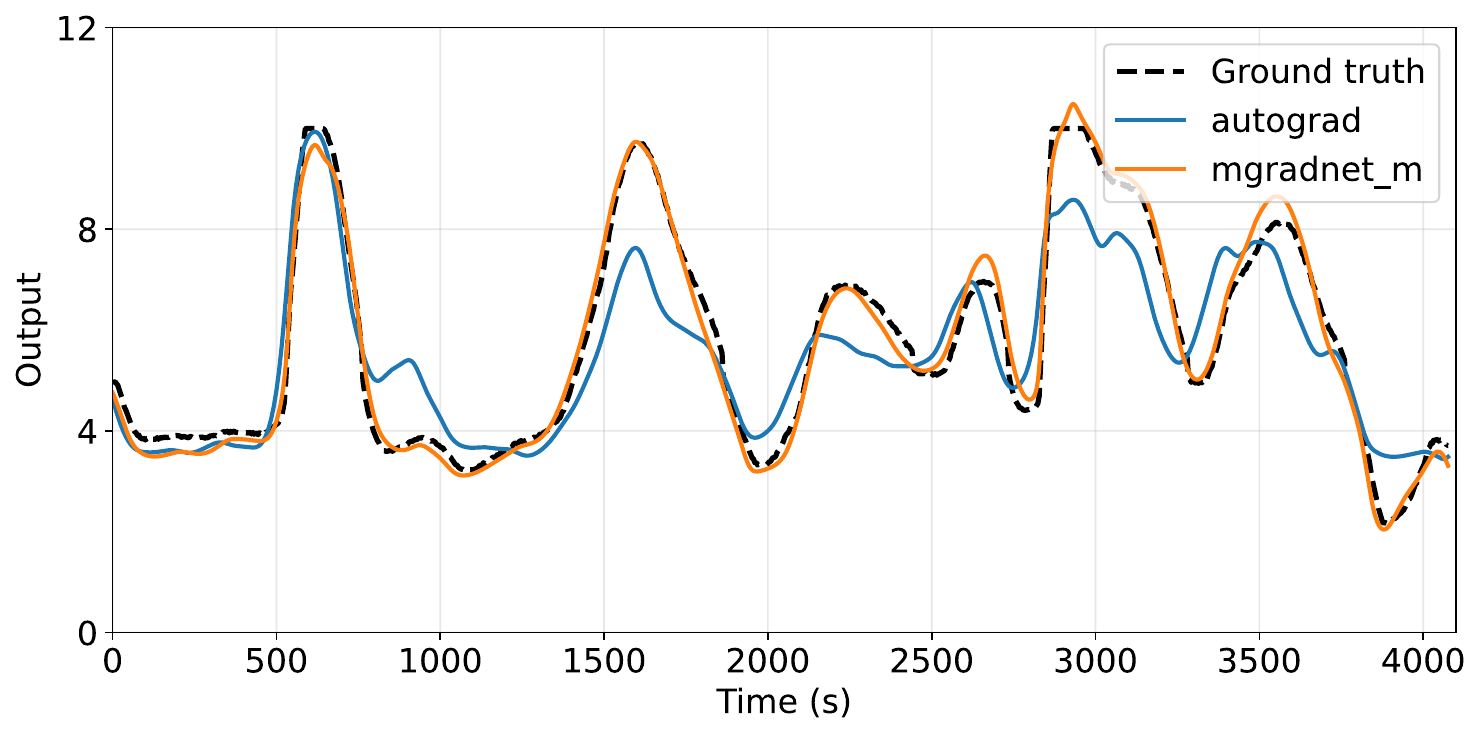}\vspace{-3mm}
    \caption{CT: test simulation over the whole test record.}
    \label{fig:CT-simulation}
\end{figure}

Figs.~\ref{fig:MSD-simulation} and \ref{fig:CT-simulation} show the test simulations of the models with the lowest validation loss. For MSD, a 100-s section of the 2000-s test record is shown. Both models follow the true output closely, but \texttt{AG-ours} deviates more near the peaks. For CT, \texttt{MGM} follows the measured level well, even during overflow, while \texttt{AG-ours} deviates from the true system.

\subsubsection{Noise level}
Table~\ref{tab:SNR-results} shows the effect of the noise level for MSD. Note that \texttt{MGM} achieves lower NRMSEs than both \texttt{AG-ref} and \texttt{AG-ours} at all SNR levels. While the best models get better as the SNR increases, the mean NRMSE remains similar across various SNR levels, indicating that the underlying dynamics are learned reliably regardless of the noise level. As the models are tested on a noise-free dataset, the model errors at $\mathrm{SNR} = 15$~dB are below the expected $\mathrm{NRMSE} \approx 0.178$ noise floor.

\begin{table}[t]
    \centering
    \caption{Model NRMSE for the MSD system at varying noise levels ($N = 20\,000$, $T = 200$).}
    \label{tab:SNR-results}
    \begin{tabular}{cccccc}
    \toprule
     & \multicolumn{1}{c}{\texttt{AG-ref}} & \multicolumn{2}{c}{\texttt{AG-ours}} & \multicolumn{2}{c}{\texttt{MGM}} \\
     \cmidrule(lr){3-4} \cmidrule(lr){5-6}
    SNR (dB) &  & Mean & Min & Mean & Min \\
    \midrule
    15 & -- & 0.097 & 0.075 & 0.062 & \textbf{0.032} \\
    35 & 0.025 & 0.104 & 0.068 & 0.061 & \textbf{0.020} \\
    40 & 0.023 & 0.109 & 0.065 & 0.059 & \textbf{0.013} \\
    50 & 0.019 & 0.097 & 0.063 & 0.063 & \textbf{0.015} \\
    \bottomrule
    \end{tabular}
\end{table}

\subsubsection{Amount of training data}
Table~\ref{tab:dataset-size} shows the effect of dataset size $N$ for MSD. Smaller datasets contain the first $N$ samples of the full training dataset, while validation and test datasets remain unchanged. As expected, access to more data reduces the error for all models. At the same time, while \texttt{MGM} outperforms \texttt{AG-ours} for all $N$, \texttt{AG-ref} obtains the best models at $N=5\,000$ and $N=10\,000$. This is likely due to the difference in implementations, indicating that the training procedure for \texttt{AG-ref} might make more efficient use of the available data.

\begin{table}[t]
    \centering
    \caption{Model NRMSE for the MSD system at training dataset sizes $N$ ($\mathrm{SNR} = 50$~dB, $T = 200$).}
    \begin{tabular}{cccccc}
    \toprule
     & \multicolumn{1}{c}{\texttt{AG-ref}} & \multicolumn{2}{c}{\texttt{AG-ours}} & \multicolumn{2}{c}{\texttt{MGM}} \\
    \cmidrule(lr){3-4} \cmidrule(lr){5-6}
    $N$ &  & Mean & Min & Mean & Min \\
    \midrule
    1\,000 & -- & 0.841 & 0.611 & 0.438 & \textbf{0.146} \\
    2\,000 & -- & 0.381 & 0.214 & 0.218 & \textbf{0.096} \\
    5\,000 & \textbf{0.054} & 0.277 & 0.094 & 0.102 & 0.077 \\
    10\,000 & \textbf{0.025} & 0.301 & 0.086 & 0.079 & 0.067 \\
    20\,000 & 0.019 & 0.097 & 0.063 & 0.063 & \textbf{0.015} \\
    \bottomrule
    \end{tabular}
    \label{tab:dataset-size}
\end{table}

\subsubsection{Simulation horizon}
Table~\ref{tab:simulation-horizon} shows the effect of the horizon $T$.
All models tend to perform better with longer horizons for the MSD system, with an exceptionally good minimum NRMSE for \texttt{MGM} at $T=100$.
Even so, \texttt{MGM} significantly outperforms the autograd competitors, achieving reasonable models at the short horizons and matching the performance of \texttt{AG-ref} at $T=200$ at a four times shorter horizon. A similar story holds for the CT benchmark, although the error obtained by \texttt{AG-ref} is the best there. Additionally, the reference uses four RK4 steps per sampling interval, compared to one for \texttt{MGM}, yielding 240 RK4 steps per data section for \texttt{AG-ref} compared to 120 for \texttt{MGM}. Consequently, \texttt{MGM} requires fewer computations for equal simulation horizons, while already matching the performance of the autograd models at shorter horizons, enabling a significantly faster training process.

\begin{table}[t]
    \centering
    \caption{Model NRMSE for both systems at horizons $T$ (MSD: $N = 20\,000$, $\mathrm{SNR} = 50$~dB).}
    \begin{tabular}{ccccccc}
    \toprule
     & & \multicolumn{1}{c}{\texttt{AG-ref}} & \multicolumn{2}{c}{\texttt{AG-ours}} & \multicolumn{2}{c}{\texttt{MGM}} \\
    \cmidrule(lr){4-5} \cmidrule(lr){6-7}
    System & $T$ &  & Mean & Min & Mean & Min \\
    \midrule
    MSD & 20 & -- & 1.000 & 0.941 & 0.572 & \textbf{0.415} \\
        & 50 & -- & 0.836 & 0.368 & 0.118 & \textbf{0.019} \\
        & 100 & 0.024 & 0.391 & 0.099 & 0.065 & \textbf{0.010} \\
        & 200 & 0.019 & 0.097 & 0.075 & 0.063 & \textbf{0.015} \\
    \midrule
    CT  & 15 & -- & 0.866 & 0.601 & 0.884 & \textbf{0.420} \\
        & 30 & -- & 0.856 & 0.761 & 0.570 & \textbf{0.236} \\
        & 60 & \textbf{0.133} & 0.743 & 0.625 & 0.273 & 0.146 \\
        & 120 & -- & 0.688 & 0.463 & 0.212 & \textbf{0.125} \\
    \bottomrule
    \end{tabular}
    \label{tab:simulation-horizon}
\end{table}

\subsubsection{Computational complexity}
\texttt{MGM} reduces the parameters for the Hamiltonian from 369 to 126 for MSD and from 69 to 28 for CT compared to \texttt{AG-ours}. Additionally, Hamiltonian-gradient inference time is decreased by 63.6\% and 50.3\% for MSD and CT, respectively. The overall training for \texttt{AG-ours} remains 1.8\% and 9.5\% faster, likely due to more efficient backpropagation through its tanh-based architecture compared with \texttt{MGM}'s nested softplus transformations and  sigmoid derivatives. By utilizing shorter simulation horizons, \texttt{MGM} can match the accuracy of \texttt{AG-ours} with a shorter learning procedure, or achieve higher accuracy when given equal training time.

\section{Conclusion}
\label{sec:conclusion}
This paper proposes an alternative parametrization for the port-Hamiltonian dynamics in OE-PHNNs. By reducing the number of parameters and omitting the automatic differentiation step, the adapted models train faster and achieve better accuracy for both simulated and benchmark systems. Furthermore, through the explicit modelling of gradients, mathematically transparent inductive biases can be incorporated into the model architecture, paving the way for future work to study the systematic selection of suitable gradient network architectures based on known properties of the underlying system.

\begin{ack}
The authors would like to thank Sarvin Moradi for her insight and assistance with this research.
\end{ack}

\end{document}